# Comparison of the Recently proposed Super Marx Generator Approach to Thermonuclear Ignition with the DT Laser Fusion-Fission Hybrid Concept by the Lawrence Livermore National Laboratory.

F. Winterberg

University of Nevada

# Abstract

The recently proposed Super Marx generator pure deuterium micro-detonation ignition concept is compared to the Lawrence Livermore National Ignition Facility (NIF) Laser DT fusion-fission hybrid concept (LiFE) [1]. In a Super Marx generator a large number of ordinary Marx generators charge up a much larger second stage ultra-high voltage Marx generator, from which for the ignition of a pure deuterium micro-explosion an intense GeV ion beam can be extracted. A typical example of the LiFE concept is a fusion gain of 30, and a fission gain of 10, making up for a total gain of 300, with about 10 times more energy released into fission as compared to fusion. This means a substantial release of fission products, as in fusion-less pure fission reactors. In the Super Marx approach for the ignition of a pure deuterium micro-detonation a gain of the same magnitude can in theory be reached [2]. If feasible, the Super Marx generator deuterium ignition approach would make lasers obsolete as a means for the ignition of thermonuclear micro-explosions.

## 1. Introduction

Since 1954 I have been actively involved in inertial confinement fusion research at a time this research was still classified in the US. I had independently discovered the basic principles and presented them in 1956 at a meeting of the Max Planck Institute in Goettingen organized by von Weizsaecker. To reach high temperatures, these principles are the Guderley convergent shock wave and the Rayleigh imploding shell solutions. The abstracts of the meeting still exist and are in the library of the University of Stuttgart.

The meeting was overwhelmed by the optimism to ignite a deuterium plasma, with the Max Planck Institute proposing a stellarator-like magnetic confinement configuration. At that time the ignition of a thermonuclear reaction with Guderley's convergent shock wave solution seemed only possible with the deuterium-tritium (DT) reaction, and for that reason was not considered worth for funding. However, following the publication of a paper by Trubnikov and Kudryavtsev [3], presented at the 1958 2$^{nd}$ United Nations Conference for the Peaceful Use for Atomic Energy, showing the importance of the electron synchrotron losses from a magnetized plasma, the hopes for a viable deuterium fusion plasma magnetic confinement configuration had been given up in favor of deuterium-tritium magnetic confinement configurations. But because a burning deuterium-tritium plasma is primarily the source of energetic neutrons, ideally suited for the fast fission of natural uranium or thorium, it was obvious to combine fusion with fission, with fusion producing neutrons and fission producing heat. Such an approach however, would not eliminate the generation of fission products as in a pure fission reactor, still posing a similar environmental nuclear waste disposal problem.

A laser DT inertial confinement fusion reactor configuration requires a high gain, typically of the order $10^3$, to make up for the poor laser efficiency. But the intense photon flash from the high gain fusion micro-explosion, entering the optical laser system with the velocity of light, can destroy the entire laser ignition apparatus, unless the laser is separated by a safe distance from the micro-explosion, which has its own technical problems.

The Super Marx generator concept for the ignition of a pure deuterium fusion reaction, if feasible, would not only bypass the fusion-fission hybrid concept, but would make the entire laser fusion approach obsolete and with it most of the rest, like the heavy ion and (ordinary) Marx generator electric pulse power approach, where the X-rays emitted from an array of exploding wires compresses and ignites a DT pellet.

## 2. Solution in Between Two Extremes

Up until now nuclear fusion by inertial confinement has only been achieved by using a fission explosive as a means (driver) for ignition. This is true not only for large thermonuclear explosive devices, like the 1952 pure deuterium Mike Test (carried out in the South Pacific with the Teller-Ulam configuration), but also for small deuterium-tritium (DT pellet) micro-explosions, (experimentally verified with a fission explosive at the Nevada Test Site by the Centurion-Halite experiment). From this experience we know that the ignition is easy with sufficiently large driver energies, but which are difficult to duplicate with lasers or electric pulse power. The problem therefore is not the configuration of the thermonuclear explosive, but the driver, be it for the ignition of pure deuterium (D) as in the Mike Test, or for the ignition of deuterium-tritium (DT), as in the Centurion-Halite experiment, because for sufficiently large driver energies the target configuration is of secondary importance.

I claim that substantially larger driver energies can be reached by a "Super Marx generator". It can be viewed as a two-stage Marx generator, where a bank of ordinary Marx generators assumes the role of a first stage. If the goal is the much more difficult ignition of a pure deuterium micro-explosion, the Super Marx generator must in addition to deliver a much larger amount of energy (compared to the energy of the most powerful lasers), also generate a magnetic field in the thermonuclear target that is strong enough to entrap the charged DD fusion products within the target. Only then is the condition for propagating thermonuclear burn fulfilled. For this to happen, a 100 MJ-1GeV-$10^7$ Ampere proton beam is needed. This is the sought after solution, in between the two extremes, shown in **Fig. 1**. It hopefully can be realized with the Super Marx [1].

## 3. From the Marx to the Super Marx

**Fig. 2** shows the circuit of an ordinary Marx generator, and **Fig. 3** that of a Super Marx. An artist's view of a mile-long Super Marx generator, charged up by about 100 ordinary Marx generators, and connected to the chamber in which the thermonuclear target is placed, is shown in **Figs. (4-6)**.

As shown in **Fig. 7**, the Super Marx is made up of a chain of co-axial capacitors with a dielectric which has to withstand the potential difference of $10^7$ Volt between the inner and outer conductor.

Following their charging up the Super Marx generator, the Marx generators are disconnected from the Super Marx. If the capacitors of the Super Marx can hold their charge long enough, this can be done by mechanical switches.

To erect the Super Marx, its capacitors C are switched into series by circular spark gap switches S. The capacitors of the Super Marx are magnetically levitated inside an evacuated tunnel, and magnetically insulated against the wall of the tunnel by an axial magnetic field $B$, generated by super conducting external magnetic field coils M. The magnetic insulation criterion requires that $B > E$, where $B$ is measures in Gauss and $E$ in electrostatic cgs units. If $B = 3\times10^4$ G for example, magnetic insulation is possible up to $E = 3\times10^4$ esu $\approx$ $10^7$ V/cm, at the limit of electron field emission. To withstand a voltage of $10^9$ Volt between the outer positively charged surface of the capacitors in series and the tunnel wall, then requires a distance somewhat more than one meter.

The capacitance of one co-axial capacitor with the inner and outer radius, $R_0$ and $R_1$ of length $l$ and filled with a dielectric of dielectric constant $\varepsilon$ is

$$C = \varepsilon \frac{l}{2\ln\left(R_1/R_0\right)} \text{ cm} \tag{1}$$

Assuming a breakdown strength of the dielectric larger than $3\times10^4$ V/cm, and a potential difference of $10^7$ Volt between the inner and outer conductor, the smallest distance of separation $d$ between both conductors has to be $d = R_1 - R_0 \cong 3\times10^2$ cm. If for example $l = 1.6\times10^3$ cm, $R_1 = l/2 = 8\times10^2$ cm, and $\varepsilon \cong 10$, one finds that $C \cong 2\times10^4$ cm. For these numbers the energy $e$ stored in the capacitor ($V = 10^7$ Volt $\cong 3\times10^4$ esu) is

$$e = (1/2)\,CV^2 \cong 10^{13}\ \text{erg} \tag{2}$$

which for the 100 capacitors of the Super Marx add up to $e \sim 10^{15}$ erg. About 10 times more energy can be stored, either if the radius of the capacitor is about 3 times larger, or with a larger dielectric constant, or with a combination of both. This means that for about 100 capacitors an energy $10^{16}$ erg = 1 GJ can be stored in the mile-long Super Marx.

Another idea proposed by Fuelling [4], is to place the ordinary Marx generators of the 1st stage inside the coaxial capacitors of the Super Marx. The advantage of this configuration is that it does not require to disconnect the Marx generators from the capacitors of the Super Max prior to its firing. Because the charging and discharging of the Super Marx can there be done very fast, one can use compact water capacitors where $\varepsilon \cong 80$. And instead of magnetic insulation of the capacitors of the Super Marx against the outer wall one can perhaps use transformer oil for the insulation. Giving each inner segment of the Super Marx enough buoyancy, for example by adding air chambers, these segments can be suspended in the transformer oil. There the outer radius of the co-axial capacitors is much larger. This permits to store in the Super Marx gigajoule energies.

If the danger of electric breakdown in the transformer oil should still persist, one might try to prevent the breakdown oil, by a rapid spiraling flow of the transformer oil, disrupting the formation of stepped leaders leading to the breakdown [5].

## 4. Connecting the Super Marx to the Load

As shown in **Fig. 8**, the last capacitor of the Super Marx is guided to the load by a cylindrical Blumlein transmission line. Because it is uncertain if a Blumlein transmission line can withstand a voltage of $10^9$ Volt long enough, one may consider a different configuration shown in **Fig. 9a**, where a superconducting capacitor containing a large toroidal magnetic field is chosen as the last capacitor. There, because of the large azimuthal magnetic field set up by the toroidal current magnetically insulates the $10^9$ Volt charged up capacitors against breakdown to the wall. This idea was previously proposed by the author in a concept to charge up the forces to gigavolt potentials by a beam of charged particles [6, 7].

The load is the deuterium target shown in **Fig. 10**. It consists of a solid deuterium rod covered with a thin ablator placed inside a cylindrical hohlraum. To the left of the hohlraum and the deuterium rod is a mini-rocket chamber filled with solid hydrogen.

The method of discharging the energy from the Blumlein transmission line to the target then goes as follows (see **Fig. 11**): 1. A short laser pulse is projected into the mini-rocket chamber, with the laser beam passing through a small hole in the center of the Blumlein transmission line. 2. By heating the hydrogen in the mini rocket chamber to a high temperature, a supersonic hydrogen jet is emitted through the Laval nozzle towards to the end of the Blumlein transmission line, forming a bridge in between the Blumlein transmission line and the target. 3. A second laser pulse then traces out an ionization trail inside the hydrogen jet, facilitating an electric discharge between the Blumlein and the target, with the space charge neutralizing plasma pinching the proton beam down to a small diameter. The large magnetic field of the $10^7$ Ampere discharge current favors the creation of a proton beam in the hydrogen rich jet, not only because of the Alfvén limit for electrons, but also because of the large radiation friction of GeV electrons going in proportion to $\gamma^2 = \left(1 - v^2/c^2\right)^{-1}$ [8]. Since for GeV protons $\gamma \approx 1$, but for electrons $\gamma \geq 10^3$, the friction force on the electrons is larger by many orders of magnitude. For a GeV -$10^7$A proton beam, focused down to an area of 0.1 cm$^2$, one computes a proton number density in the beam equal to $n_b \cong 2 \times 10^{16}$ cm$^{-3}$. To have a non-relativistic electron return current in the plasma of the hydrogen jet, the number density of the jet must be $n_e \geq n_b$. This condition should be satisfied, since the jet comes from the dense laser heated solid hydrogen in the mini rocket chamber attached to the target. If for the discharge to the target instead of the Blumlein transmission line the magnetically insulated superconducting torus, shown in **Fig. 9a**, is used, the charging and discharging of the torus is done from a spherical electrode in the center of the torus as shown in **Fig 9b**.

## 5. Thermonuclear Ignition and Burn

For the deuterium-tritium thermonuclear reaction the condition for propagating burn in a sphere of radius *r* and density *ρ*, heated to a temperature of $10^8$ K, is given by $\rho r \geq 1$ g/cm$^2$. This requires an energy of about 1 MJ. For the deuterium reaction this condition is $\rho r \geq 10$ g/cm$^2$, with an ignition temperature about 10 times larger. That a thermonuclear detonation in deuterium is possible at all is due to the secondary combustion of the T and $He^3$ DD fusion reaction products [9]. The energy required there would be about $10^4$ times larger or about $10^4$ MJ, for all practical purposes out of reach for non-fission ignition. However, if the ignition and burn is along a deuterium cylinder, where the charged fusion products are entrapped within the cylinder, the condition $\rho r \geq 10$ g/cm$^2$ can be replaced by

$$\rho z \geq 10 \text{ g/cm}^2 \tag{3}$$

where *z* is the length of the cylinder. The entrapment is possible if a large current is flowing through the cylinder, generating a large azimuthal magnetic field. The condition to entrap the charged fusion products by the azimuthal magnetic field is there given by the inequality

$$r_f < r_c \tag{4}$$

In (4) $r_f$ is the charged fusion product Larmor radius, and $r_c$ the radius of the deuterium cylinder, where

$$r_f = \frac{Mvc}{ZeB} \tag{5}$$

with *B* the magnetic field strength, *M* and *Z* the mass and charge, and *v* the velocity of the charged fusion products. By order of magnitude one has $v \sim c/10$. A current *I* [A] flowing through the deuterium cylinder produces a magnetic field which at the surface of the cylinder of radius $r_c$ is

$$B = \frac{0.2I}{r_c} \tag{6}$$

Inserting (5) and (6) into (4) one has

$$I > \frac{5Mvc}{Ze} \tag{7}$$

This inequality is well satisfied if $I \geq 10^7$ A.

If the GeV -$10^7$ Ampere proton beam passes through background hydrogen plasma with a particle number density $n$, it induces in the plasma a return current carried by its electrons, where the electrons move in the same direction as the protons. But because the current of the proton beam and the return current of the plasma electrons are in opposite directions, they repel each other. Since the stagnation pressure of the GeV -$10^7$ Ampere proton beam is much larger than the stagnation pressure of the electron return current, the return current electrons will be repelled from the proton beam, towards the surface of the proton beam.

The stagnation pressure of a GeV proton beam is ($M_H$ proton mass)

$$p_i \cong \rho_i c^2 = n_b M_H c^2 \tag{8}$$

For $n_b \cong 2\times10^6$ cm$^{-3}$ one obtains $p_i \cong 3\times10^{13}$ dyn/cm$^3$.

For the electron return current one has ($m$ electron mass)

$$p_e = n_e m v^2 \tag{9}$$

With the return current condition $n_e e v_e = n_i e v_i$, where for GeV protons $v_i \cong c$, one has

$$v_e / c = n_i / n_e \tag{10}$$

Taking the value $n_e = 5\times10^{22}$ cm$^{-3}$, valid for uncompressed solid deuterium, one obtains $v_e \cong 10^4$ cm/s and hence $p_e \cong 5\times10^3$ dyn/cm$^2$. This is negligible against $p_i$, even if $n_e$ is $10^3$ times larger, as in highly compressed deuterium. The assumption, that the magnetic field of the proton beam is sufficiently strong to entrap the charged fusion products within the deuterium cylinder, is therefore well justified.

If the charged fusion products are entrapped within the deuterium cylinder, and if the condition $\rho z > 10$ gm/cm$^2$ is satisfied, and finally, if the beam energy is large enough that a length $z$ > ($10/\rho$) cm is heated to a temperature of $10^9$ K, a thermonuclear detonation wave can propagate down the cylinder. This then leads to large fusion gains.

The stopping length of single GeV protons in dense deuterium is much too large to fulfill inequality (3). But this is different for an intense beam of protons, where the stopping length is determined by the electrostatic proton-deuteron two-stream instability [10]. In the presence of a strong azimuthal magnetic field the stopping length is enhanced by the formation of a collision less shock [11]. For the two-stream instability alone, the stopping length is given by

$$\lambda \cong \frac{1.4c}{\varepsilon^{1/3}\omega_i} \tag{11}$$

where $c$ is the velocity of light, and $\omega_i$ the proton ion plasma frequency, furthermore $\varepsilon = n_b/n$, with $n$ the deuterium target number density. For a 100-fold compressed deuterium rod one has $n = 5\times10^{24}\,\text{cm}^{-3}$ with $\omega_i = 2\times10^{15}\,\text{s}^{-1}$. One there finds that $\varepsilon = 4\times10^{-9}$ and $\lambda \cong 1.2\times10^{-2}\,\text{cm}$. This short length, together with the formation of the collision-less magneto-hydrodynamic shock, ensures the dissipation of the beam energy into a small volume at the end of the deuterium rod. For a deuterium number density $n = 5\times10^{24}\,\text{cm}^{-3}$ one has $\rho = 17\,\text{g/cm}^3$, and to have $\rho z > 10$ g/cm$^2$, then requires that $z \geq 0.6\,\text{cm}$. With $\lambda < z$, the condition for the ignition of a thermonuclear detonation wave is satisfied.

The ignition energy is given by

$$E_{ign} \sim 3nkT\pi r^2 z \tag{12}$$

where $T \approx 10^9$ K.

For 100-fold compressed deuterium, one has $\pi r^2 = 10^{-3}$ cm$^2$, when initially it was $\pi r^2 = 10^{-1}$ cm$^2$. With $\pi r^2 = 10^{-3}\,\text{cm}^2$, $z = 0.6\,\text{cm}$ one finds that $E_{ign} \leq 10^{16}\,\text{erg}$ or $\leq$ 1GJ. This energy is provided by the $10^7$ Ampere-GeV proton beam lasting $10^{-7}$s. The time is short enough to ensure the cold compression of deuterium to high densities. For a $10^3$-fold compression, found feasible in laser fusion experiments, the ignition energy is ten times less.

In hitting the target a fraction of the proton beam energy is dissipated into X-rays by entering and bombarding the high $Z$ material cone, focusing the proton beam onto the deuterium cylinder. The X-rays released fill the hohlraum surrounding the deuterium cylinder, compressing it to high densities, while the bulk of the proton beam energy heats and ignites the deuterium cylinder at its end, launching in it a detonation wave.

Both the energy and the magnetic field are supplied by the proton beam from the Super Marx generator, more than what all other inertial confinement fusion drivers are capable.

## 6. Conversion of the explosively released energy

As shown in **Fig. 6**, the deuterium micro-explosion takes place inside an evacuated cavity. If this cavity has the radius $R$ and if it is filled with a magnetic field of strength $B$, it contains the magnetic energy

$$e_M = \frac{4\pi}{3} R^3 \frac{B^2}{8\pi} = \frac{1}{6} R^3 B^2 \tag{13}$$

Let us assume that $R$ = 15 meter $= 1.5\times10^3$ cm, and that $B = 2\times10^4$ G (which can be reached with ordinary electromagnets), one finds that $e_M = 23$ GJ.

The rapidly expanding fully ionized fire ball of the deuterium micro-explosion pushes the magnetic field towards the cavity wall, and if the wall is covered with induction coils, the released fusion energy is converted into an electromagnetic pulse lasting for the time $\tau \cong R/a$, where $a \cong 10^8$ cm/s is the expansion velocity of the fireball. For $R = 1.5\times10^3$ cm, one finds $\tau \cong 1.5\times10^{-5}$ s. This time is long enough for the pulse to be stretched out and converted into a useful electromagnetic energy.

For an ignition an energy of 100 MJ and a yield of 23 GJ, the fusion gain would be $G$ = 230, about the same as for the LiFE concept. However, since even in pure deuterium burn neutrons are released through the secondary combustion of the tritium D-D fusion reaction products, a much higher overall gain is possible with an additional fission burn, as in the LiFE concept [1].

## 7. Other possibilities

The energy of up to a gigajoule, delivered in ~ $10^{-7}$ seconds at a power of $10^{16}$ Watt, opens up other interesting possibilities.

1. If instead of protons heavy ions are accelerated with such a machine at gigavolt potentials, these ions will upon impact be stripped off of all their electrons, in case of uranium all of its 92 electrons. Accordingly this would result in a 92 fold increase of the beam current to ~ $10^9$ Ampere. With such an ultrahigh current, a very different fusion target, shown in **Fig. 12**, seems possible, where a solid deuterium rod is placed inside a hollow metallic cylinder. The inner part of the beam $I_1$ will directly pass through the deuterium inside the cylinder, while the outer part of the beam $I_0$ will be stopped in the cylindrical shell, there depositing its energy and imploding the shell onto the deuterium cylinder, at the same time compressing the azimuthal magnetic beam field inside the cylinder. If ignited at the position where the beam hits and ignites the cylinder, will lead to a deuterium detonation wave propagating down the cylinder.

2. At a beam current of ~ $10^9$ Ampere, will lead to a large, inward directed magnetic pressure. At a beam radius of 0.1 cm, the magnetic field will be of the order of $2\times10^9$ Gauss, with a magnetic pressure of $10^{17}$ dyn/cm$^2$ ≈ $10^{11}$ atmospheres. At these high pressures the critical mass of fissile material (U235, Pu 23a, and U233) can be reduced to ~ $10^{-2}$ g [12-14]. This would make possible micro-fission explosion reactors not having the meltdown problem of conventional fission reactors.

3. In general, the attainable very high pressures would have many interesting applications. One example is the release of fusion energy from exotic nuclear reactions, like the $pB^{11}$ neutron-free fusion reaction, conceivably possible under very high pressures.

## 8. Discussion

The fusion/fission LiFE concept proposed by the Livermore National Laboratory is an outgrowth of the DT laser ignition project pursued at the National Ignition Facility. Ignition is there expected in the near future. With its large fission component, it is difficult to see how the LiFE can compete with conventional fission reactors. Like them it still has the fission product nuclear waste problem. For this reason it hardly can without fission solve the national energy crisis, for what it has been billed by California Governor Schwarzenegger [15].

The proposed Super-Marx concept is by comparison a much more ambitious project, because it recognizes that the fundamental problem of inertial confinement fusion is the driver energy, not the target. And that only with order of magnitude larger driver energies can real success be expected. This in particular is true, if the goal is to burn deuterium. Unlike deuterium which is everywhere abundantly available, the burn of deuterium-tritium depends on the availability of lithium, a comparatively rare element. In conclusion, we display a table comparing the two different concepts.

It is also worthwhile to compare the Super Marx generator approach for thermonuclear micro-explosion ignition with the work by Basko and his group [16, 17]. They hope to achieve the same with heavy ion particle accelerators. And they too propose cylindrical targets, with magnetic fields (both axial and azimuthal), to entrap the charged fusion products. They can make a good case for the ignition of DT microexplosions, much better than what is possible with lasers, but because it is difficult to reach petawatt megajoule energies with a space charge limited particle accelerator, they think the ignition of the deuterium reaction is possible only with the help of a DT "tablet", very much as in my 1982 concept [9]. In the absence of such a DT ignitor, their calculations predict beam energies larger than 100MJ for compression and ignition, with comparable gains and yields as in the Super Marx generator approach. They share the conclusion, that the ignition of thermonuclear micro-explosions, most likely requires much larger energies than those hoped for, and more in line what I had thought to be attainable with a levitated superconducting capacitor back in 1968 [6], with the prospect of gigajoule energies released in about $10^{-7}$ seconds.

**Acknowledgement**

This work has been supported in part by the US Department of Energy under Grant No. DE-FG02-06ER54900. The author thanks Dr. S. Fuelling for the many valuable comments and the artwork of the many drawings.

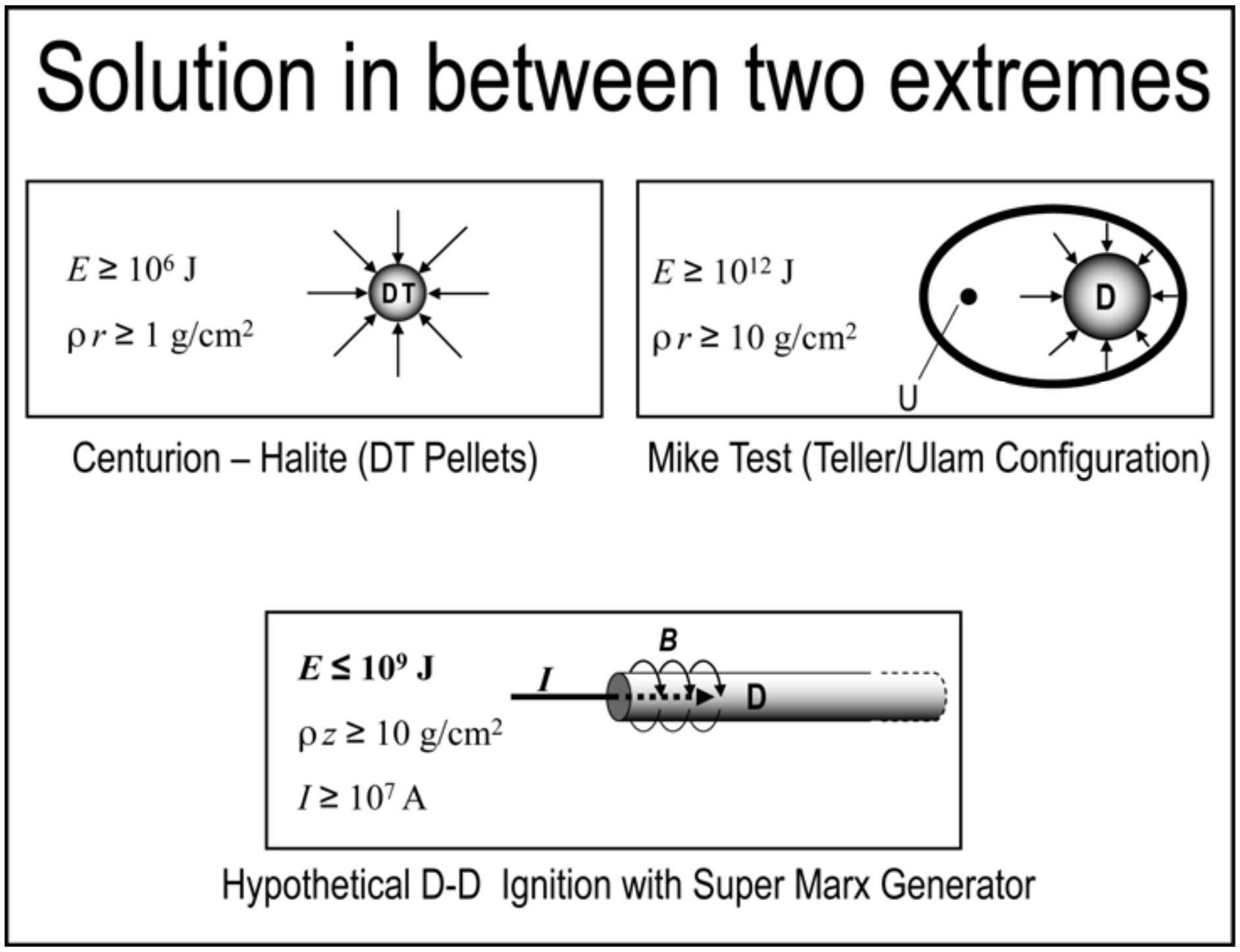

Figure 1: Ignition of a deuterium target by a GeV-10 MA proton beam

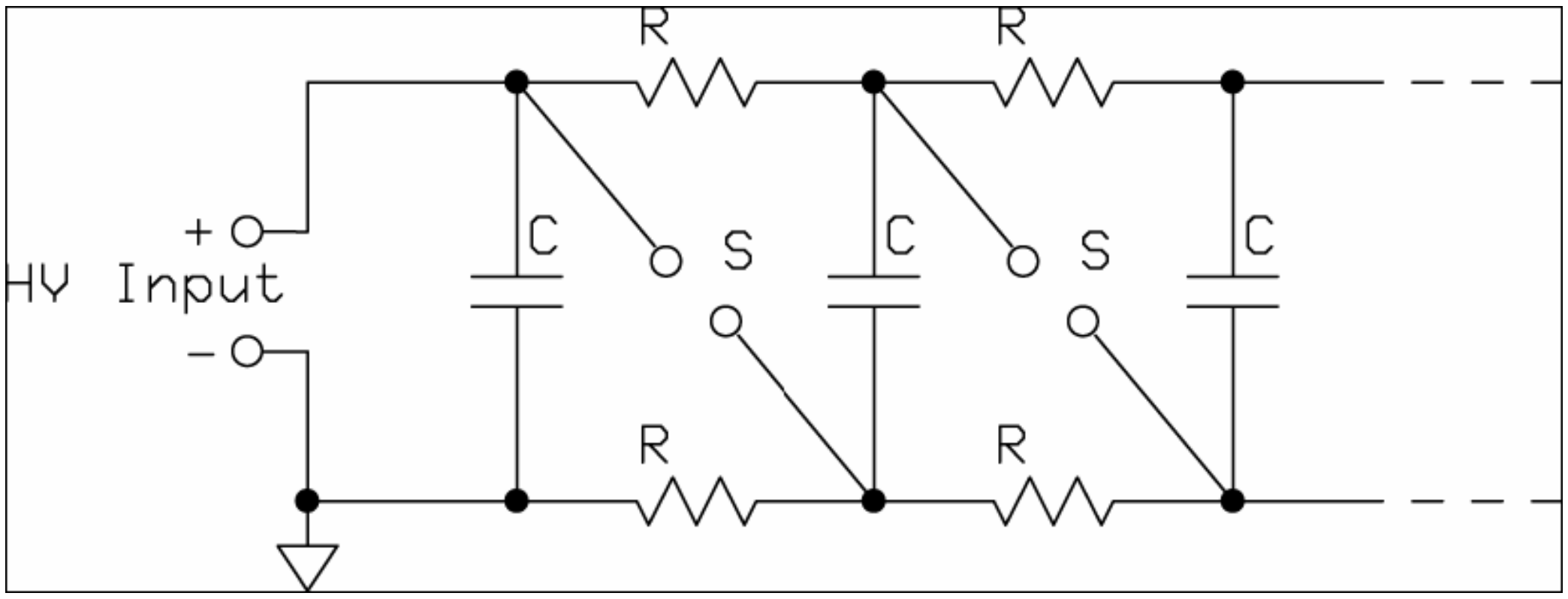

Figure 2: In an “ordinary” Marx generator n capacitors $C$ charged up to the voltage $v$, and are over spark gaps switched into series, adding up their voltages to the voltage $V = nv$.

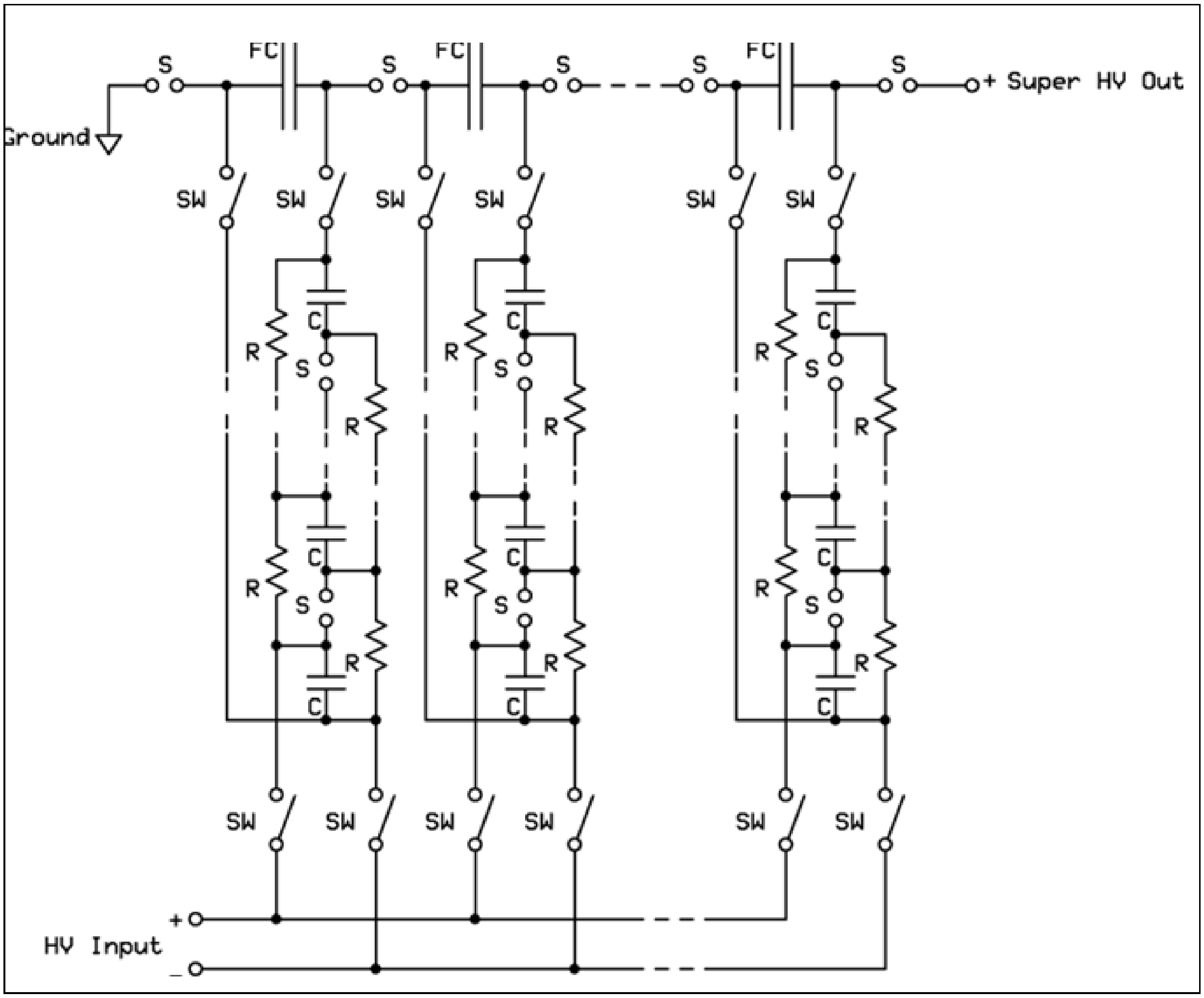

Figure 3: In a Super Marx generator, *N* Marx generators charge up *N* fast capacitors FC to the voltage *V*, which switched into series add up their voltages to the voltage *NV*.

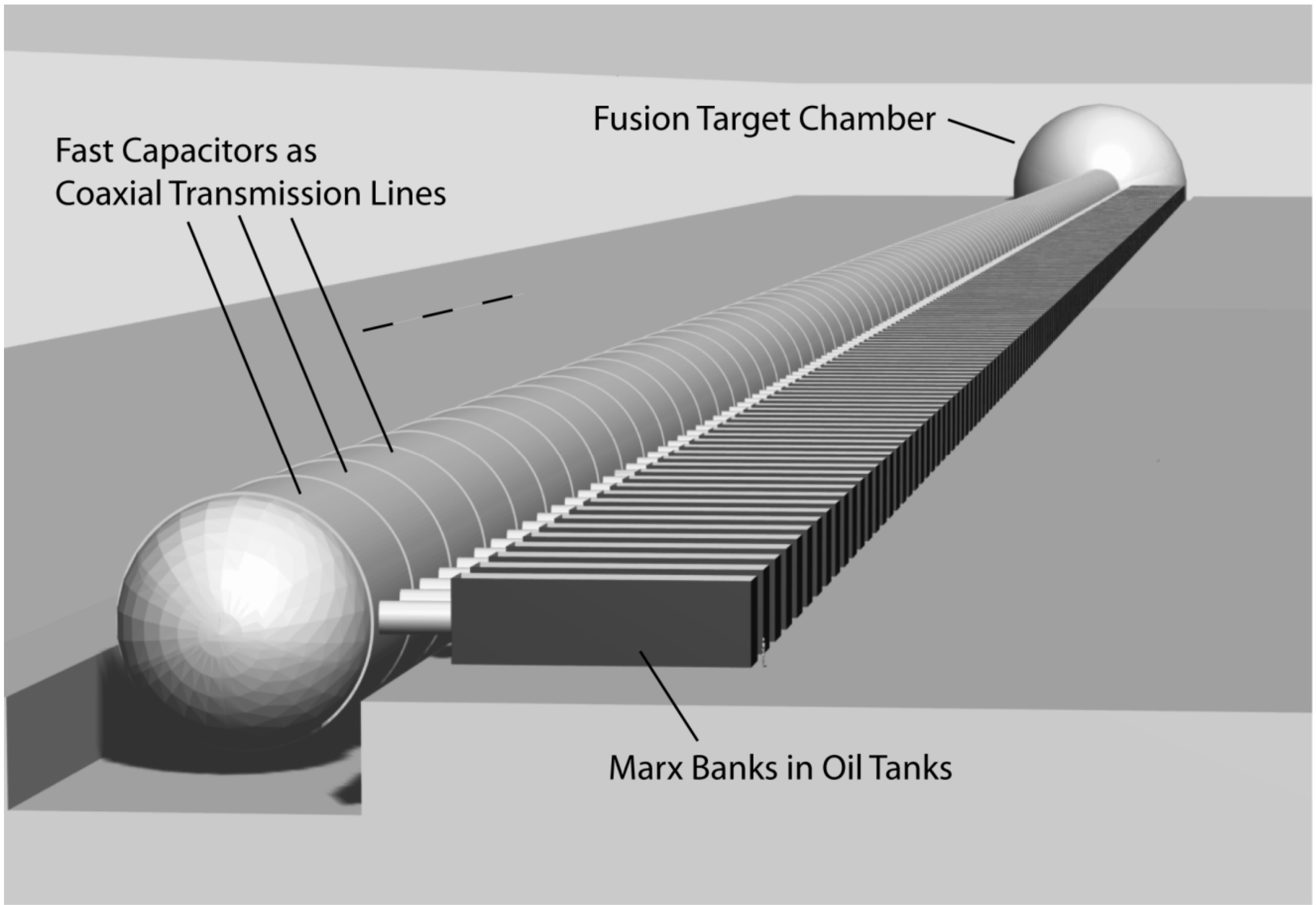

Figure 4: Artistic perception of a 1.5 km long Super Marx generator, composed of 100x 15 m long high voltage capacitors each designed as a magnetically insulated coaxial transmission line. The coaxial capacitors/transmission lines are placed inside a large vacuum vessel. Each capacitor/transmission line is charged by two conventional Marx generators up symmetrically to 10 MV (± 5 MV). After charge-up is completed, the Marx generators are electrically decoupled from the capacitors/transmission lines. The individual capacitors/transmission lines are subsequently connected in series via spark gap switches (i.e. the ‘Super Marx’ generator), producing a potential of 1 GV.

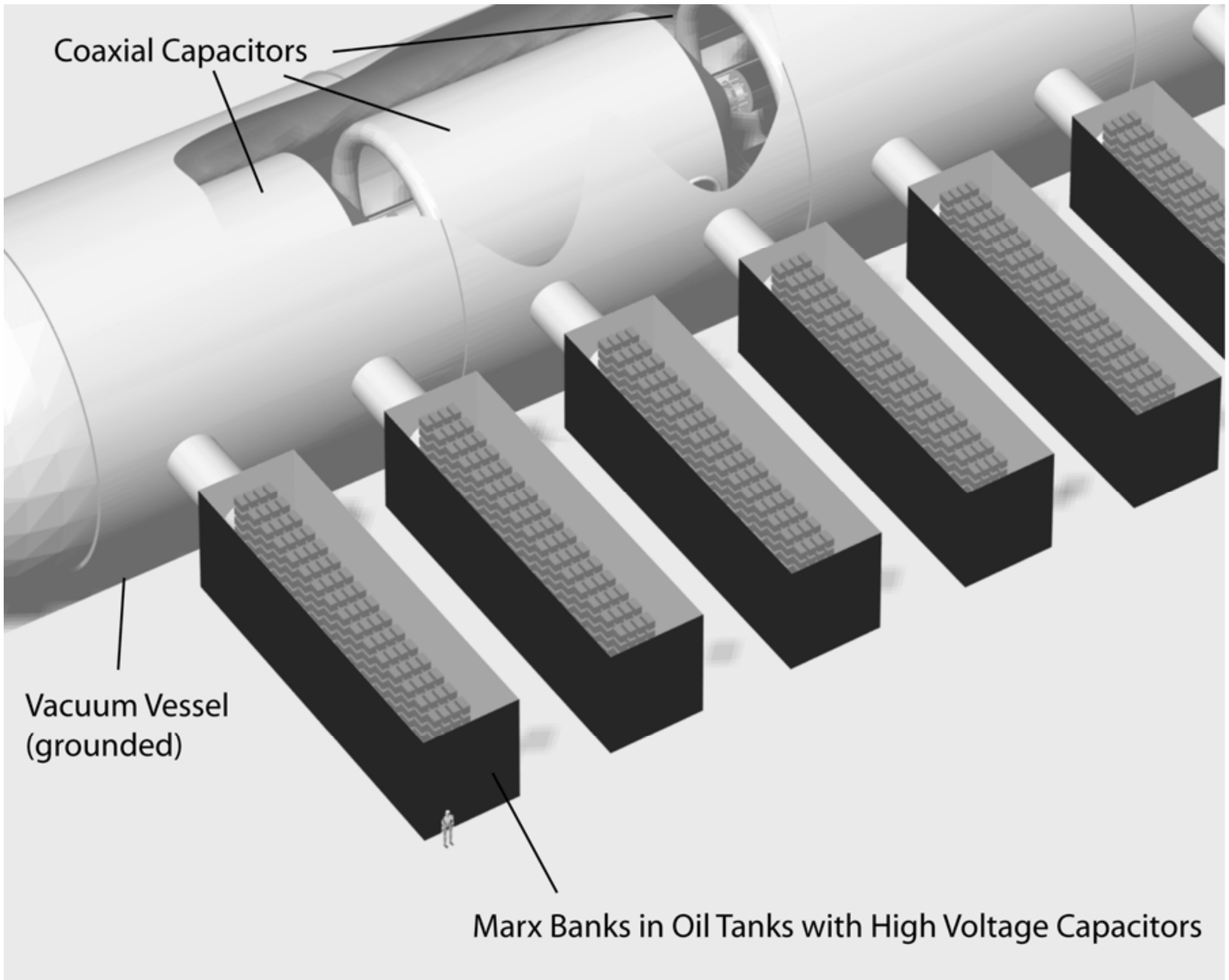

Figure 5:  Detail view of a section of the Super Marx generator. Two conventional Marx banks charge up one coaxial capacitor/transmission line element to 10 MV.

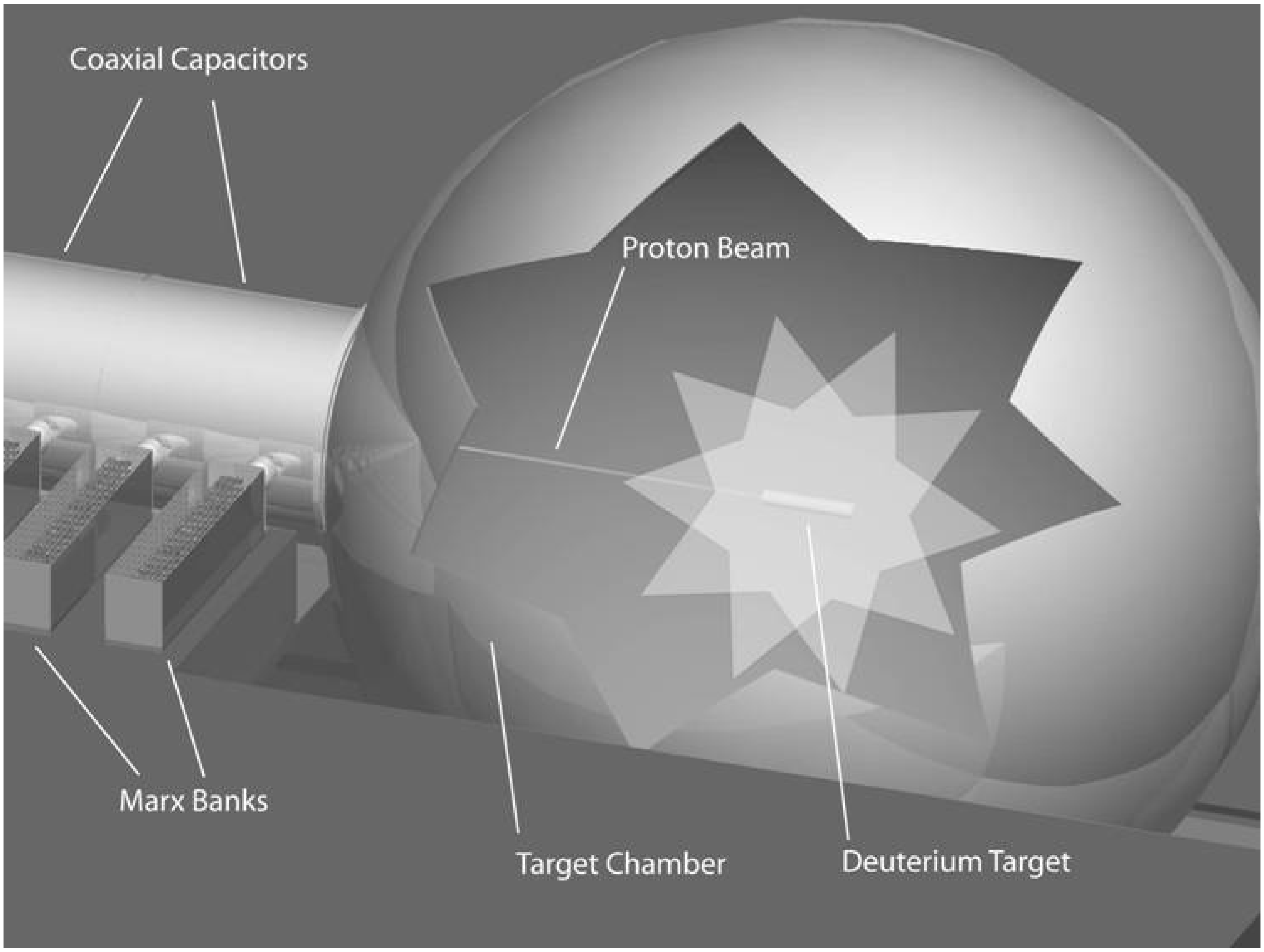

Figure 6: Injection of GeV – 10 MA proton beam, drawn from Super Marx generator made up of magnetically insulated coaxial capacitors into chamber with cylindrical deuterium target.

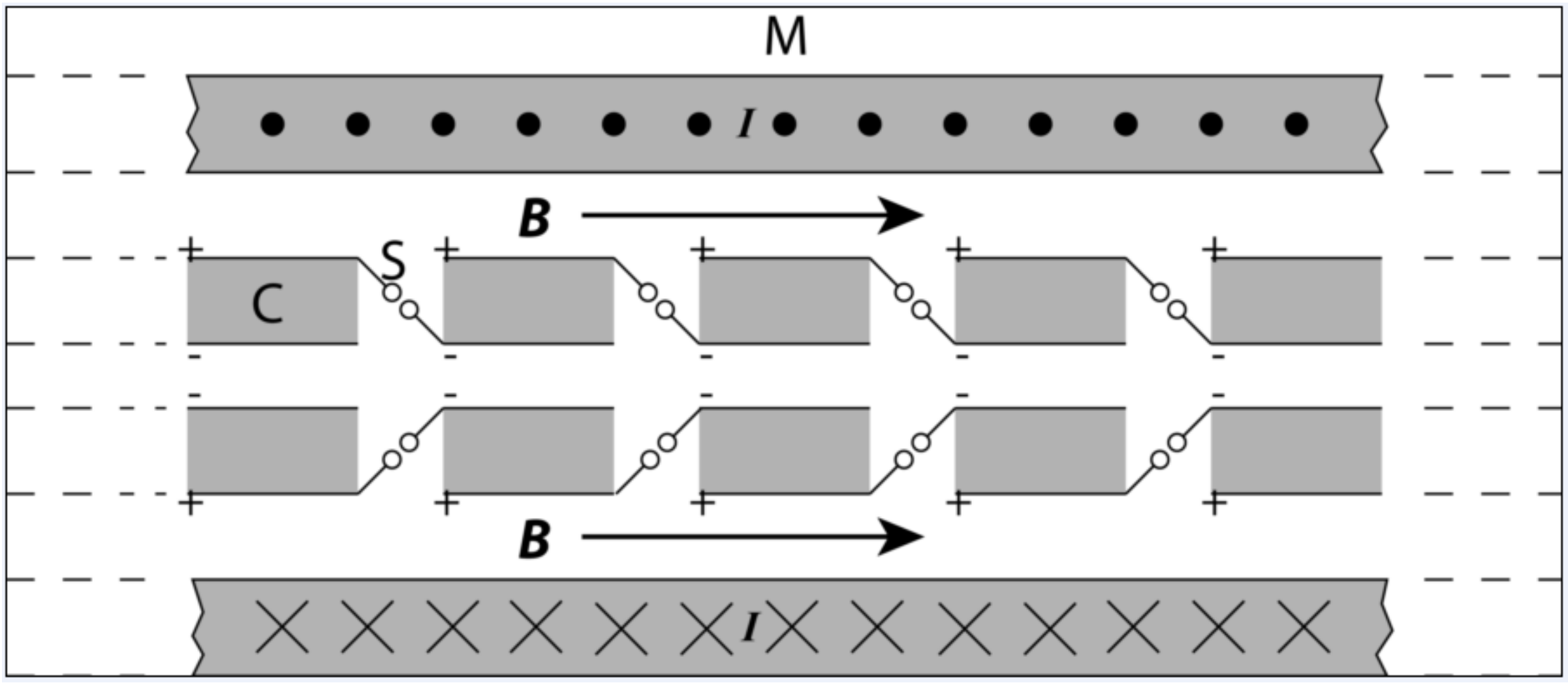

Figure 7: Showing a few elements of the Super Marx generator.

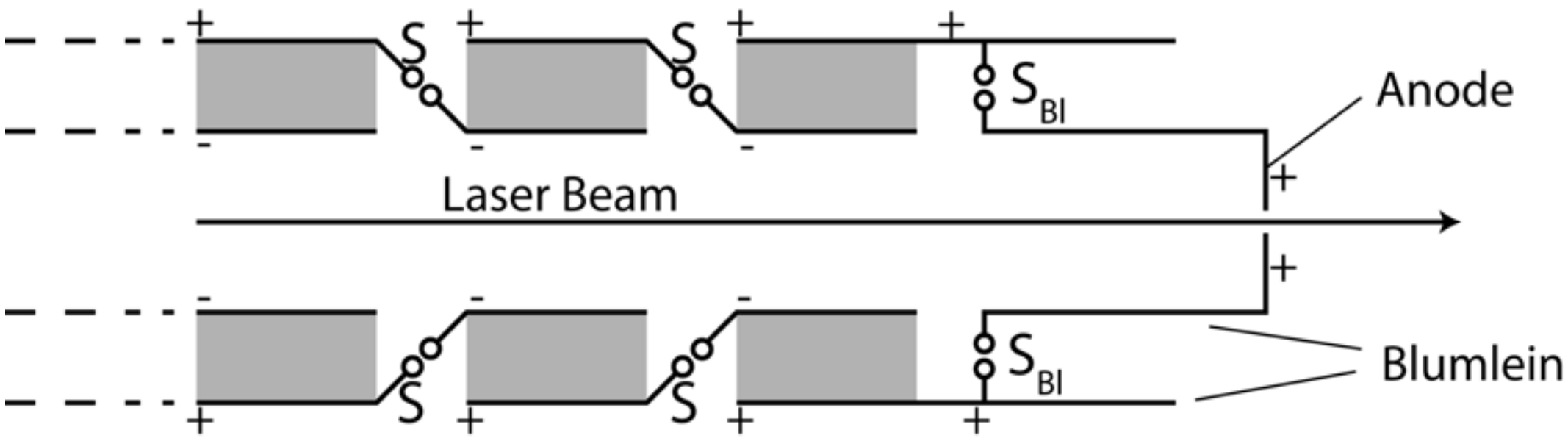

Figure 8: Connection of the last capacitor of the Super Marx to the Blumlein transmission line.

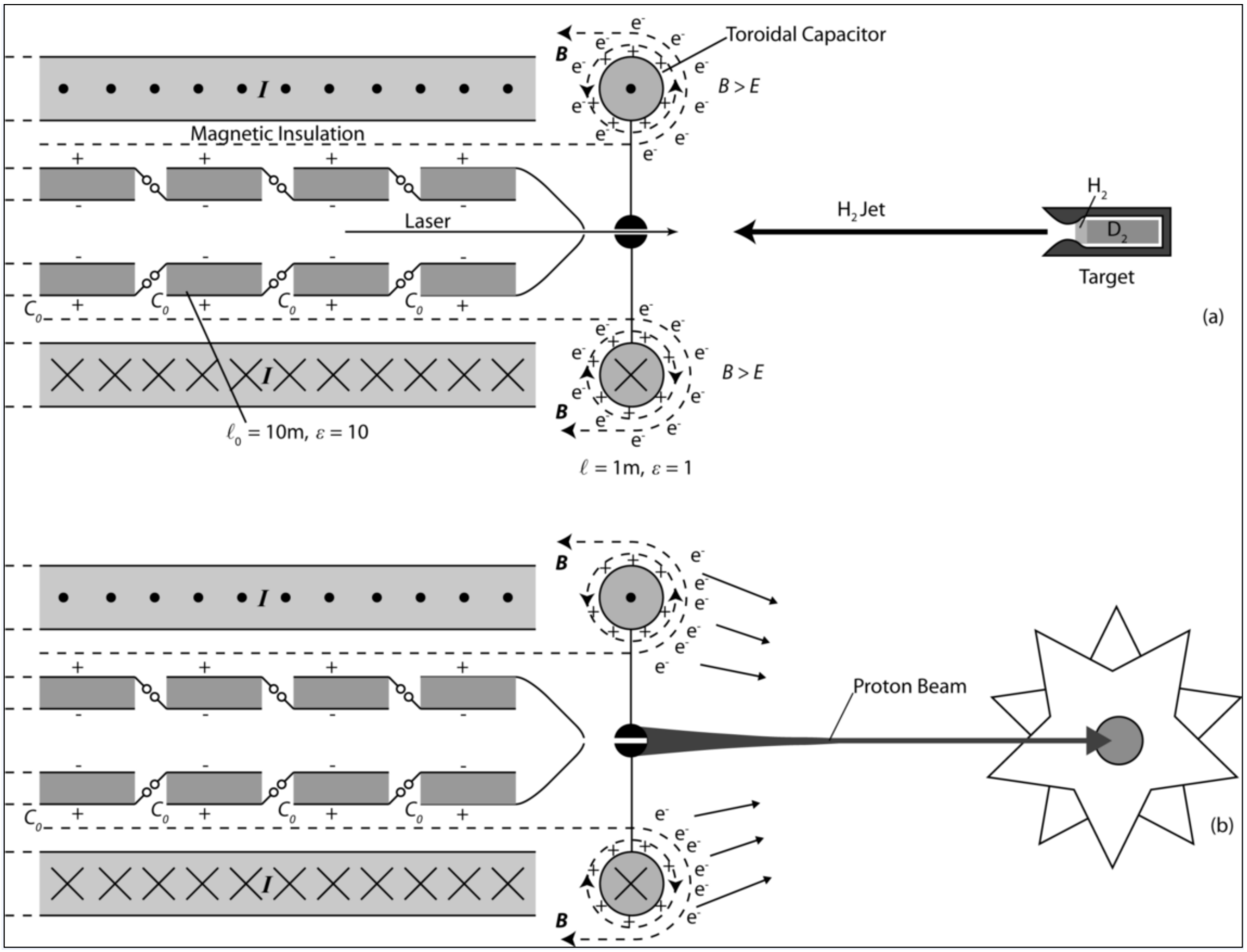

Figure 9: The superconducting toroidal capacitor (a) and its discharge onto the target (b).

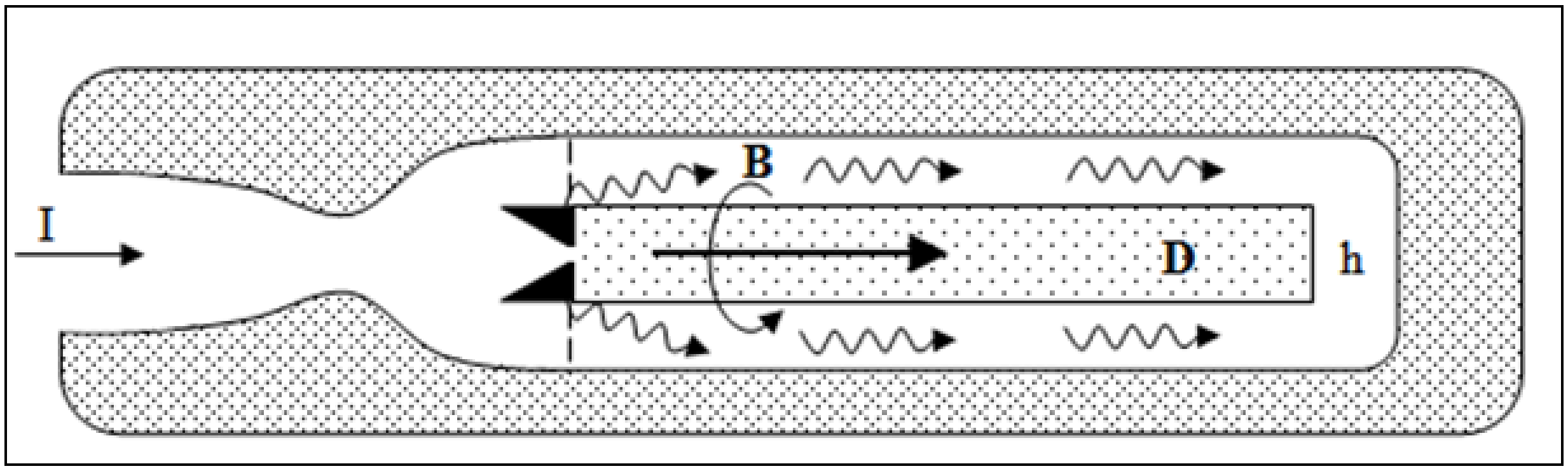

Figure 10: Possible deuterium micro-detonation target: I ion beam, D deuterium cyinder, B magnetic field, h cylindrical hohlraum

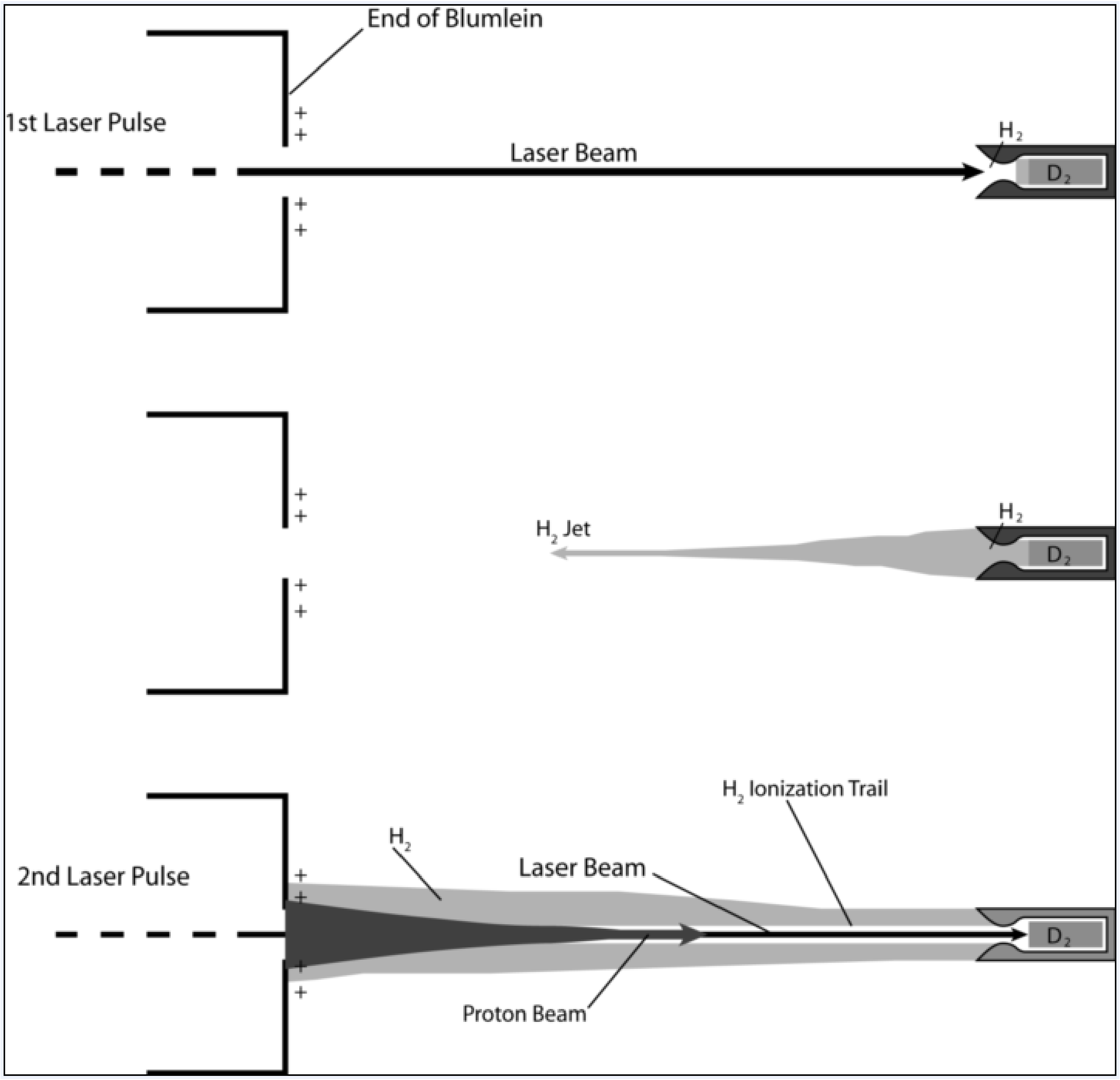

Figure 11: Sequence of events to bombard the target by the proton beam from the Blumlein transmission line

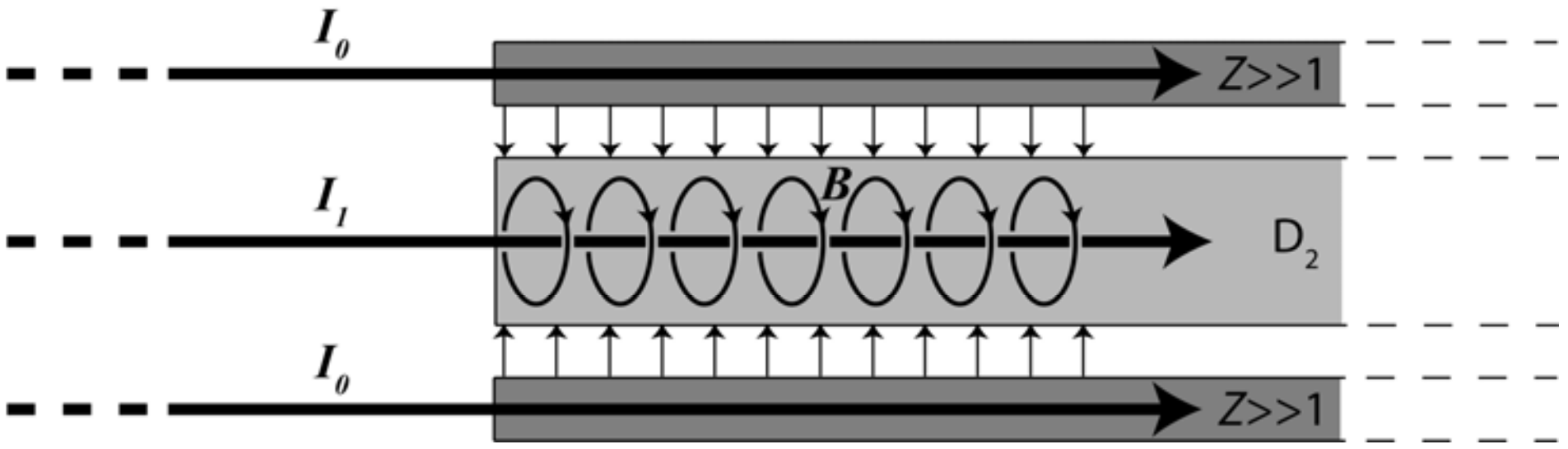

Figure 12: Bombarding a cylindrical, deuterium containing target, with an intense heavy ion beam.